\documentclass[prl,twocolumn]{revtex4}%
\usepackage{amsmath}
\usepackage{amsfonts}
\usepackage{amssymb}
\usepackage{graphicx}%
\begin{document}
\title{Group theoretical description of artificial magnetic
metamaterials utilized for negative index of refraction}
\author{W. J. Padilla}
\email{willie@lanl.gov} \affiliation{Los Alamos National
Laboratory, MS G756, MST-CINT, Los Alamos, NM 87545.}

\begin{abstract}
Group theoretical methods are used to determine the
electromagnetic properties of artificial magnetic meta-materials,
based solely upon the symmetries of the underlying constituent
particles. Point groups for such materials are determined. From
the transformation properties of an electromagnetic (EM) basis
under symmetries of the particles, it is possible to determine,
(i) the EM modes of the particles, (ii) the form of constitutive
relations (iii) magneto-optical response of a meta-material or
lack thereof. These methods are shown to be useful for
determination of the isotropic or bi-anisotropic nature of
artificial magnetic particles. The results for several artificial
magnetic metamaterials are given. Based upon these methods, we
predict an ideal planar artificial meta-material, which eliminates
an undesirable electric resonance while still exhibiting a
magnetic response. Further we determine the subset of point groups
of which particles must belong to in order to yield an isotropic
3D magnetic response, and we show an example.

\end{abstract}

\maketitle

Progress in the rapidly advancing field of metamaterials has
experienced even further growth due to the recent observation of a
negative index (NI) medium.\cite{smith1}  A NI medium is an
artificial material in which both the magnetic $\mu$ and electric
$\epsilon$ response obtain simultaneous negative values, thus
yielding a negative index of refraction. Although predicted over
three decades ago,\cite{veselago} it wasn't until recently that
such materials were realized. Typically NI media utilize
individual components for the electric and magnetic response.
Materials which exhibited negative epsilon, known as artificial
dielectrics, have been known since the
40's.\cite{kock,bracewell,rotman} The critical component
responsible for the demonstration of NI was the realization that
artificial non-magnetic materials could be constructed to exhibit
negative response.\cite{pendry1}

The first demonstrations of NI materials were performed at
microwave frequencies, due in part to the ease of fabrication as
well as simplicity of measurements. \cite{smith1,shelby} The
magnetic component of NI materials have since been demonstrated at
lower RF\cite{wiltshire} frequencies and higher THz\cite{padilla}
frequencies. Great interest for the further extension of these
exotic materials to optical frequencies (utilizing nano-sized
elements) is fuelling massive research efforts. Given the
potential of NI materials to span the electromagnetic spectrum, it
is important to understand their full complex electromagnetic
behavior. In particular, the structures utilized for NI are
bianisotropic and may yield rich electromagnetic properties such
as chirality $\kappa$ and/or non-reciprocity $\chi$.

At microwave frequencies where complex reflection and transmission
measurements (S-parameter) are common, it is difficult to
characterize bianisotropic materials. This difficulty stems from
the fact that these materials are necessarily described by the
most general form of the constitutive relations. There may be 36
complex quantities to determine, and thus standard complex
reflection and transmission measurements yield incomplete EM
information. At THz and higher frequencies where phase sensitive
measurements are not common it is even more difficult, although
methods such as ellipsometry or THz time domain spectroscopy may
prove useful. Further since a magnetic and electric resonant
response enter similarly into the Fresnel equations, it is
difficult to determine the origin of the response. Typically,
artificial magnetic metamaterials are constructed from conducting
elements, and thus a full electromagnetic characterization is a
necessity, since an electric response is unavoidable. Analytical
theories have yielded equations able to predict the resonant
frequency $\omega_{0}$ and plasma frequency $\omega_{p}$ of
metamaterials. However there is a lack of suitable analytical
methods capable of determining the many varying and complicated EM
properties. Simulation is still heavily relied upon and, as of
yet, is still unable to determine and distinguishing bianisotropic
response.

\begin{figure}
[ptb]
\begin{center}
\includegraphics[
width=3.25in,keepaspectratio=true
]%
{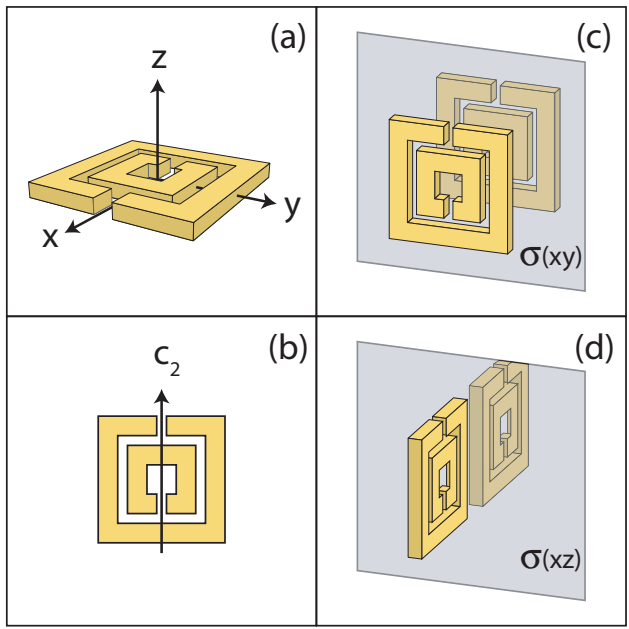}%
\caption{Point group symmetries of the SRR particle. Panel (a)
shows the coordinate system convention. Panel (b) shows the
symmetry axis of the SRR and the C$_{2}$ symmetry (rotation about
the axis by 2$\pi$/n, n=2). Panels (c) and (d) demonstrate the
mirror plane
symmetries.}%
\label{fig1}%
\end{center}
\end{figure}

We present a group theoretical method capable of determining
various electromagnetic properties for magnetic metamaterials.
This method is based simply upon the symmetry operations of the
constituent particles about a point in space, i.e. point group
theory. An EM basis is assigned to the artificial particle under
investigation, and transformations of this basis under the
symmetries of the group yield the electromagnetic modes. This new
method allows one to calculate: the EM modes, the form of the
constitutive relations, and the determine wether a particle will
exhibit magneto-optical response. This approach is demonstrated by
way of an example.

Materials utilized for negative magnetic response have been shown
to be bianisotropic,\cite{marques} thus let us review the
constitutive relations for these materials, which can be written
as:\cite{kong}
\begin{equation}\label{eq1}
 \left[ \begin {array}{c} \overline{D}\\\noalign{\medskip}
 \overline{B}\end {array} \right]
 = \left[ \begin {array}{cc} \overline{\overline{\epsilon}}
 &\overline{\overline{\xi}}\\\noalign{\medskip}
 \overline{\overline{\zeta}}&\overline{\overline{\mu}}\end {array}
  \right]  \left[ \begin {array}{c} \overline{E}\\\noalign
  {\medskip}\overline{H}\end {array}
   \right]
\end{equation}

In Eq.\ref{eq1} the permittivities within the 2x2 matrix are
tensors of second rank, also called dyadics. The terms
$\overline{\overline{\xi}}$ and $\overline{\overline{\zeta}}$ are
called the magneto-optical permittivities, and they describe
coupling of the magnetic to electric response and electric to
magnetic response respectively.

As an example, let us start by examining the point group
symmetries of a typical element utilized for magnetic response. In
Fig. \ref{fig1} (b-d) the symmetry operations of a split ring
resonator (SRR) are shown. A symmetry operation brings the element
into self coincidence. Thus for the SRR it can be seen that there
are three symmetries which meet this criterion. In addition, all
groups also contain the identity operation E. We then find that
the SRR particle belongs to the C$_{2v}$ point group which
contains the following elements [E, c$_{2}$, $\sigma(xy)$,
$\sigma(xz)$] where we use the Schoenflies notation.

\begin{figure*}
[ptb]
\begin{center}
\includegraphics[
width=6.5in,keepaspectratio=true
]%
{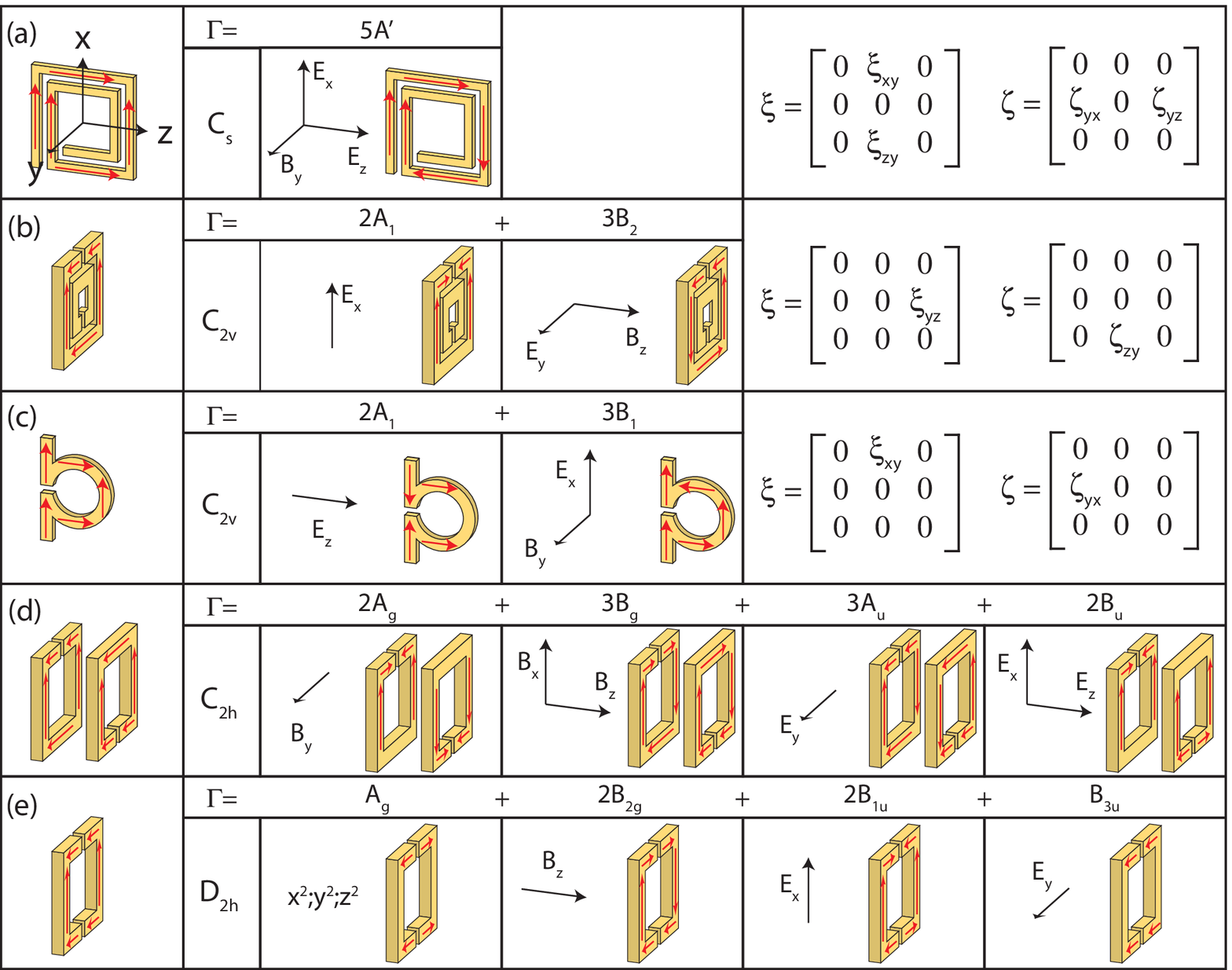}%
\caption{Basis utilized (red arrows in left column) for magnetic
metamaterials used to calculate the EM modes (red arrows in the
remaining columns). The remaining columns for each row show the
SALCs and modes of the SRR particle, as determined by point group
symmetries. For each column, the irrep is shown above and the
corresponding
component of the an external electromagnetic wave, or function is shown.}%
\label{fig2}%
\end{center}
\end{figure*}

We now turn to determination of the EM modes of the particles.
Since we have determined SRRs to belong to the C$_{2v}$ group we
can utilize the character table for this analysis(see Table I).
With the character table we can assign a basis set to the SRR and
see how this basis set transforms under the symmetry operations of
the SRR. For brevity we only consider the outer split ring,
however the inner ring is also easily handled by this method. We
want our basis to represent areas of electrical activity, thus we
choose the basis shown in the left column of Fig. \ref{fig2}(b).
Regions marked with arrows represent areas which can be polarized
by an external electric field (i.e. currents can flow in these
directions). These are similar to the P orbitals used in molecular
orbital group theory (MOGT). The next step is to write out
matrices which describe how this basis transforms under C$_{2v}$.
For example, there are five vectors which make up our basis for
the SRR, and since the identity leaves the particle unchanged,
this would be a 5x5 matrix with 1's along the diagonal. Once we
obtain matrix representations for each element of the group, we
can use a result of the Orthogonality Theorem to determine how
many times each irreducible representation (irrep) occurs. For our
chosen basis we use the following equation,\cite{kettle}
\begin{equation}\label{eq8}
 a_m={\frac {1}{h}}\sum_c{n_c\chi(g)\chi_m(g)}
\end{equation}

\noindent where $h$ is the order of the group (h=4 in this case),
$n_c$ is the number of symmetry operation in each class, $\chi(g)$
are the characters of the original representation, and $\chi_m$
are the characters of the m$^{th}$ irreducible representation.
Using Equation (\ref{eq8}) we find our basis is spanned by
$\Gamma_{SRR}$=2A$_1$+3B$_2$.

\begin{table}[hbtp]
\centering
\begin{tabular}{|c|c c c c|c|c|}
\hline
{C$_{2v}$} & E & C$_2$ & $\sigma(xz)$ & $\sigma(xy)$ & Linear & Quadratic   \\
\hline
A$_1$    & 1 &    1  &     1      &      1   & x & x$^2$,y$^2$,z$^2$  \\
A$_2$    & 1 &    1  &    -1      &     -1   & R$_x$ & yz  \\
B$_1$    & 1 &   -1  &     1      &     -1   & z,R$_y$ & xz  \\
B$_2$    & 1 &   -1  &    -1      &      1   & y,R$_z$ & xy  \\
\hline
$\Gamma_{SRR}$ & 5 &   -1  &    -1       &      5   &      &      \\
\hline
\end{tabular}
\label{table2}\caption{Character table for the C$_{2v}$ point
group. The body of the table lists the characters, (trace of a
matrix representation) of the group. The first column lists the
irreps of the group, the top of columns 2-5 lists the symmetry
operations, and the last 2 columns lists a number of linear and
quadratic functions that transform as the various irreps of the
group. The bottom row lists the characters of our chosen SRR
basis.}
\end{table}

Once we know the irreps spanned by an arbitrary basis set, we can
work out the appropriate linear combinations of basis functions
that transform the matrix representations of our original
representation into block diagonal form. These are called symmetry
adapted linear combinations (SALCs). We use a projection operator
to determine the SALCs, that transforms as an irrep. This is given
by,
\begin{equation}\label{eq9}
 \phi_i'=\sum_g{\chi_{k}(g)g\phi_i}
\end{equation}

\noindent where  $\phi_i'$ is the SALC,  $\chi_{\kappa}$ is the
character of the k$^{th}$ irrep, $g$ is the symmetry operation,
and $\phi_i$ is the basis function.

We can normalize the SALCs determined by Equation (\ref{eq9}), but
this is not necessary since a constant factor doesn't affect the
symmetry of the calculated modes. The response of the SRRs can now
be determined by considering incident external electromagnetic
fields. For example by examination of the character table for the
C$_{2v}$ point group we see that light polarized along the
$\hat{x}$-axis transforms as A$_1$, since it transforms in the
same manner as the function x. Thus $\hat{y}$-polarized light
transforms as B$_2$ symmetry. The function R$_{\alpha}$ represents
rotation about the $\alpha$ axis, where
$\alpha$=$\hat{x},\hat{y},\hat{z}$. Thus a magnetic field
polarized along the $\hat{z}$-direction also transforms as B$_2$
symmetry. We summarize these results in Fig. \ref{fig2} for: (a)
planar spirals, (b) SRRs, (c) Omega particles\cite{engheta} (d) an
SRR and its enantiomer\cite{enantiomer} (e) symmetric ring
resonator.

An electric field polarized along the $\hat{x}$-axis (A$_{1}$) of
the SRR drives currents as shown in row (b) of Fig. \ref{fig2}.
This would give a frequency response determined by the dimensions
(length) of the SRR segments along which E$_x$ lies. For
$\hat{y}$-polarized (B$_2$) light much more exotic behavior is
predicted. Notice that for the irrep of B$_2$ both y and R$_z$
form a suitable basis. Thus we can use a linear combination of
these two functions for the basis. This predicts that the SRR
should exhibit a magneto-optical response, as in accord with
MOGT.\cite{kettle} Furthermore, since E$_y$ and B$_z$ are the same
basis, they will occur at the same frequency. In other words, an E
field polarized along the $\hat{y}$-axis will result in a resonant
response at a frequency $\omega_0$, and a magnetic field polarized
along the $\hat{z}$-axis will result in a resonant response at the
same frequency $\omega_0$. These theoretical predictions are
consistent with results obtained from a simple analytical
model\cite{marques} as well as by simulation.\cite{katsarakis}

In Fig. \ref{fig2} we show predictions for other various magnetic
metamaterials. Particles with the lowest symmetry are listed on
the top and higher symmetry particles on the bottom. The
artificial magnetic metamaterials listed in the first 3 rows are
shown to be bianisotropic, and thus we list the form of the
constitutive equations governing the predicted magneto-optical
response on the right side of each row. In row (d) we show a
particular way of symmetrizing the SRR particle, which we predict
will eliminate the magneto-optical response. The material is a
bipartite lattice with each of the two sub-lattices consisting of
SRRs each with the gap oriented oppositely. Thus it is predicted
that any polarization rotation or mixing resulting from one unit
cell is corrected by the other unit cell, resulting in no net
polarization rotation. Indeed it can be seen that the theory
predicts no magneto-optical activity, while still exhibiting a
magnetic response for B$_z$. Another predicted way to eliminate
$\xi$ and $\zeta$ is to add a second gap in the SRR opposite to
the first, as shown in Fig. \ref{fig2}(e). Again the theory
predicts no magneto-optical response and thus we have eliminated
bianisotropy by symmetrizing the SRR. This structure is simpler
than that depicted in Fig. \ref{fig2}(d) and we have a predicted
magnetic response for B$_z$.

\begin{figure}
[ptb]
\begin{center}
\includegraphics[
width=3.25in,keepaspectratio=true
]%
{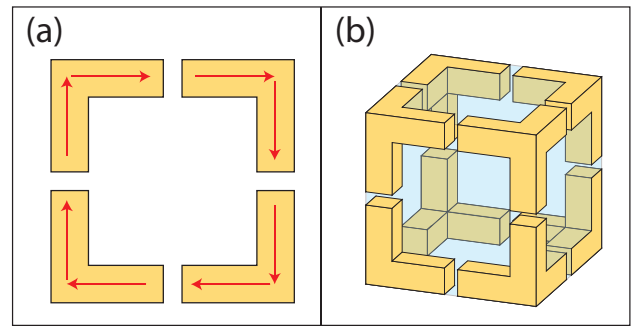}%
\caption{Predicted ideal planar and ideal 3D magnetic particles.
In panel (a) we show a planar magnetic particle with D$_{4h}$
symmetry. The currents under the A$_{2g}$ magnetic mode are shown.
Panel (b) shows a 3D isotropic magnetic particle with O$_h$
symmetry. The electric response of this particles is also
isotropic, but importantly does not occur at the same frequency as
the magnetic resonance.}%
\label{fig3}%
\end{center}
\end{figure}

Lastly let us turn to the prediction of an ideal magnetic
metamaterial. With our new understanding of the irreps of a point
group and their relation to magneto-optical behavior we can
leverage point group tables for help. An ideal magnetic particle
is one in which no magneto optical behavior is predicted. In the
parlance of group theory this implies we should look for a point
group which has linear basis functions with rotational functions
R$_\alpha$ but with little or no occurrences of linear
$\hat{x},\hat{y},\hat{z}$ functions. This not only ensures the
elimination of the $\xi$ and $\zeta$ terms, but also off-diagonal
terms in the $\mu$ and $\epsilon$ response functions, like those
which occur for the particle listed in Fig. \ref{fig2}(d), i.e.
3B$_g$ and 2B$_u$. In Table II we show a candidate point group
which should have good magnetic response with no magneto-optical
activity, and no frequency dependent $\epsilon$ occurring near the
magnetic resonance. A particle which has the symmetry of this
group is shown if Fig. \ref{fig3}.

Group theoretical analysis is carried out for the particle
depicted in Fig. \ref{fig3}(a), and we find the following modes
$\Gamma$=A$_{1g}$+A$_{2g}$+B$_{1g}$+B$_{2g}$+2E$_u$. Thus the only
linear modes determined are a magnetic mode (A$_{2g}$) and an
electric mode (E$_u$). The fact that the electric mode does not
occur in the same irrep as the magnetic mode ensures it will not
occur at the same frequency. Further the two dimensional E$_u$
mode implies the electric response of the particle will be
identical along the $\hat{x}$ and $\hat{y}$ directions with no
cross coupling terms ($\epsilon_{xy}$=$\epsilon_{yx}$=0). Thus if
we want to construct a 3D \textit{isotropic} magnetic metamaterial
free from magneto-optical activity, then the geometry of the
constituent particles are required to be one of the following
point groups: T$_h$,T$_d$,I$_h$, and O$_h$. The two simplest of
which to visualize is I$_h$, an icosahedron, and O$_h$, a cube
(depicted in Fig. \ref{fig3}(b)) or octahedron.
\begin{table}[hbtp]
\centering
\begin{tabular}{|c|c c c c c c c c c c|c|}
\hline
{D$_{4h}$} & E & 2C$_4(z)$ & C$_2$& 2C'$_2$& 2C''$_2$&i&2S$_4$&$\sigma_h$ & 2$\sigma_v$&2$\sigma_d$ & Linear \\
\hline
A$_{1g}$    & 1 & 1 & 1 & 1 & 1 & 1 & 1 & 1 & 1 & 1 & -  \\
A$_{2g}$    & 1 & 1 & 1 & -1 & -1 & 1 & 1 & 1 & -1 & -1 & R$_z$  \\
B$_{1g}$    & 1 & -1 & 1 & 1 & -1 & 1 & -1 & 1 & 1 & -1 & -  \\
B$_{2g}$    & 1 & -1 & 1 & -1 & 1 & 1 & -1 & 1 & -1 & 1 & -  \\
E$_g$    & 2 & 0 & -2 & 0 & 0 & 2 & 0 & -2 & 0 & 0 & (R$_x$,R$_y$)  \\
A$_{1u}$    & 1 & 1 & 1 & 1 & 1 & -1 & -1 & -1 & -1 & -1 & -  \\
A$_{2u}$    & 1 & 1 & 1 & -1 & -1 & -1 & -1 & -1 & 1 & 1 & z  \\
B$_{1u}$    & 1 & -1 & 1 & 1 & -1 & -1 & 1 & -1 & -1 & 1 & -  \\
B$_{2u}$    & 1 & -1 & 1 & -1 & 1 & -1 & 1 & -1 & 1 & -1 & -  \\
E$_u$    & 2 & 0 & -2 & 0 & 0 & -2 & 0 & 2 & 0 & 0 & (x,y)  \\
\hline
\end{tabular}
\label{table3}\caption{Character table for the D$_{4h}$ point
group.}
\end{table}

We have demonstrated a new method capable of determining various
properties of the electromagnetic response for artificial magnetic
metamaterials. A specific example has been worked out for the most
common element utilized for negative magnetic response. This
analysis has also been carried out and detailed for other various
artificial magnetic metamaterials. We have predicted 2 ideal
magnetic particles, including one for 3D isotropic magnetic
response. Whether one wants to take advantage of the exotic
electromagnetic properties that emerge from the bianisotropic
nature of artificial metamaterials, or construct artificial
materials free from this complication, these novel methods are
valuable for determining the expected response.

We would like to acknowledge support from the Los Alamos National
Laboratory LDRD Director's Fellowship, and thank David Schurig,
Rick Averitt, and David Smith for valuable feedback.


\begin{thebibliography}{99}
\bibitem{enantiomer}Enantiomer in this sense is defined as ``the exact
opposite" meaning that the polarization mixing which results from
one orientation of the split gap can be corrected by another SRR
with the split gap oriented oppositely.
\end{thebibliography}
\end{document}